%% file: ssbse19_final.tex
\documentclass[a4paper,runningheads]{llncs}
\usepackage{butterma}
\idline{The final authenticated version is available online at Springer/SpringerLink.}

\usepackage[utf8]{inputenc}
\usepackage{graphicx}
\usepackage[hidelinks]{hyperref}
\hypersetup{
	bookmarks=true,         
	unicode=false,          
	pdftoolbar=true,        
	pdfmenubar=true,        
	pdffitwindow=false,     
	pdftitle={Does Diversity Improve the Test Suite Generation for Mobile Applications?},
	pdfauthor={Thomas Vogel, Chinh Tran, and Lars Grunske},
	pdfsubject={},
	pdfcreator={},
	pdfproducer={},
	pdfkeywords={Fitness Landscape Analysis, Diversity, Test Generation},
	pdfnewwindow=true,
}

\usepackage{xspace}
\usepackage{paralist}
\usepackage{cite}

\usepackage{booktabs}
\usepackage{multirow}
\usepackage{wrapfig}
\usepackage[font=small,labelfont=bf]{caption}

\usepackage{subfigure}
\usepackage{dblfloatfix}
\usepackage{xcolor}

\usepackage[]{amsmath}
\usepackage{amssymb}
\usepackage{algorithm}
\usepackage[noend]{algpseudocode}
\algnewcommand\algorithmicinput{\textbf{Input:}}
\algnewcommand\Input{\item[\algorithmicinput]}
\algnewcommand\algorithmicoutput{\textbf{Output:}}
\algnewcommand\Output{\item[\algorithmicoutput]}
\algnewcommand{\LeftComment}[1]{\Statex \(\triangleright\) #1}

\newcommand{\Sapienz}{\textnormal{\textsc{Sapienz}}\xspace}
\newcommand{\SapienzDiv}{\textnormal{\textsc{Sapienz$^{div}$}}\xspace}
\newcommand{\SapienzShort}{\textbf{S}\xspace}
\newcommand{\SapienzDivShort}{\textbf{Sd}\xspace}

\newcommand{\nsgaTwo}{\mbox{NSGA-II}\xspace}

\newcommand{\ie}{\textit{i.e.}\xspace}
\newcommand{\eg}{\textit{e.g.}\xspace}
\newcommand{\cf}{\textit{cf.}\xspace}
\newcommand{\etal}{\textit{et\,al.}\xspace}

\newcommand{\aarddict}{{aarddict}\xspace}
\newcommand{\hotdeath}{{hotdeath}\xspace}
\newcommand{\kNineMail}{{k9mail}\xspace}
\newcommand{\munchLife}{{MunchLife}\xspace}
\newcommand{\passwordmanager}{{passwordmanager}\xspace}

\begin{document}
\pagestyle{headings}
	
\title{Does Diversity Improve the Test Suite Generation for Mobile Applications?} 

\author{Thomas Vogel \and Chinh Tran \and Lars Grunske}
\authorrunning{T. Vogel \and C. Tran \and L. Grunske}

\institute{\textit{Software Engineering Group}, \textit{Humboldt-Universit\"{a}t zu Berlin}, Berlin, Germany
\email{\{thomas.vogel, grunske\}@informatik.hu-berlin.de}, \email{mail@chinhtran.de}
}

\maketitle
\thispagestyle{electronic}
\begin{abstract}
In search-based software engineering we often use popular heuristics with default configurations, which typically lead to suboptimal results, or we perform experiments to identify configurations on a trial-and-error basis, which may lead to better results for a specific problem. To obtain better results while avoiding trial-and-error experiments, a fitness landscape analysis is helpful in understanding the search problem, and making an informed decision about the heuristics.
In this paper, we investigate the search problem of test suite generation for mobile applications (apps) using \Sapienz whose heuristic is a default NSGA-II. We analyze the fitness landscape of \Sapienz with respect to genotypic diversity and use the gained insights to adapt the heuristic of \Sapienz. These adaptations result in \SapienzDiv that aims for preserving the diversity of test suites during the search. To evaluate \SapienzDiv, we perform a head-to-head comparison with \Sapienz on 76 open-source apps.
\end{abstract}

\section{Introduction}

In search-based software engineering and particularly search-based testing, popular heuristics (\eg,\cite{Mao+2016}) with best-practice configurations in terms of operators and parameters (\eg,\cite{Fraser13A}) are often used. As this out-of-the-box usage typically leads to suboptimal results, costly trial-and-error experiments are performed to find a suitable configuration for a given problem, which leads to better results~\cite{Arcuri2013}.
To obtain better results while avoiding trial-and-error experiments, fitness landscape analysis can be used~\cite{Malan+Engelbrecht2013,Pitzer+Affenzeller2012}.
The goal is to analytically understand the search problem, determine difficulties of the problem, and identify suitable configurations of heuristics that can cope with these difficulties (\cf~\cite{Malan+Engelbrecht2013,Moser+2017}).

In this paper, we investigate the search problem of test suite generation for mobile applications (apps). We rely on \Sapienz that uses a default \nsgaTwo to generate test suite for apps~\cite{Mao+2016}. \nsgaTwo has been selected as it ``is a widely-used multiobjective evolutionary search algorithm, popular in SBSE research''~\cite[p.\,97]{Mao+2016}, but without adapting it to the specific problem (instance).
Thus, our goal is to analyze the fitness landscape of \Sapienz and use the insights for adapting the heuristic of \Sapienz. This should eventually yield better test results.

Our analysis focuses on the global topology of the landscape, especially how solutions (test suites) are spread in the search space and evolve over time.
Thus, we are interested in the genotypic diversity of solutions, which is considered important for evolutionary search~\cite{Crepinsek2013}.
According to our analysis, \Sapienz lacks diversity of solutions so that we extend it to \SapienzDiv that integrates four diversity promoting mechanisms. 
Therefore, our contributions are the descriptive study analyzing the fitness landscape of \Sapienz (Section~\ref{sec:fla-sapienz}), \SapienzDiv (Section~\ref{sec:sapienzdiv}), and the empirical study with 76 apps evaluating \SapienzDiv (Section~\ref{sec:evaluation}).

\section{Background: \Sapienz and Fitness Landscape Analysis}
\label{sec:sapienz}\label{sec:fla}

\Sapienz is a multi-objective search-based testing approach~\cite{Mao+2016}. Using \nsgaTwo, it automatically generates test suites for end-to-end testing of Android apps.
A test suite~$t$ consists of $m$ test cases
$\left\langle s_1, s_2,...,s_m \right\rangle$,
each of which is a sequence of up to $n$~GUI-level events
$\left\langle e_1, e_2,...,e_n \right\rangle$
that exercise the app under test.
The generation is guided by three objectives:
\begin{inparaenum}[(i)]
    \item maximize fault revelation,
    \item maximize coverage, and
    \item minimize test sequence length.
\end{inparaenum}
Having no oracle, \Sapienz considers a crash of the app caused by a test as a fault.
Coverage is measured at the code (statement coverage) or activity level (skin coverage).
Given these objectives, the fitness function is the triple of the number of crashes found, coverage, and sequence length.
To evaluate the fitness of a test suite, \Sapienz executes the suite on the app under test deployed on an Android device or emulator.

A \textit{fitness landscape analysis} can be used to better understand a search problem~\cite{Malan+Engelbrecht2013}. A fitness landscape is defined by three elements (\cf~\cite{Stadler2002}):
\begin{inparaenum}[(1)]
\item A search space as a set $X$ of potential solutions.
\item A fitness function $f_k : X \rightarrow {\rm I\!R}$ for each of the $k$ objectives.
\item A neighborhood relation $N : X \rightarrow 2^{X}$ that associates neighbor solutions to each solution (\eg, using basic operators, or distances of solutions).
\end{inparaenum}
Based on these three elements, various metrics have been proposed to analyze the landscape~\cite{Malan+Engelbrecht2013,Pitzer+Affenzeller2012}. They characterize the landscape, for instance, in terms of
the global topology (\ie, how solutions and the fitness are distributed),
local structure (\ie, ruggedness and smoothness), and
evolvability (\ie, the ability to produce fitter solutions).
The goal of analyzing the landscape is to determine difficulties of a search problem and identify suitable configurations of search algorithms that can cope with these difficulties (\cf~\cite{Malan+Engelbrecht2013,Moser+2017}).

\section{Fitness Landscape Analysis of \Sapienz}
\label{sec:fla-sapienz}

\subsection{Fitness Landscape of \Sapienz}

At first, we define the three elements of a fitness landscape (\cf~Section~\ref{sec:fla}) for \Sapienz:
\begin{inparaenum}[(1)]
	\item The search space is given by all possible test suites $t$ according to the representation of test suites in Section~\ref{sec:sapienz}.
	\item The fitness function is given by the triple of the number of crashes found, coverage, and test sequence length (\cf~Section~\ref{sec:sapienz}).
	\item As the neighborhood relation we define a genotypic distance metric for two test suites (see Algorithm~\ref{alg:distance}). The distance of two test suites $t_1$ and $t_2$ is the sum of the distances between their ordered test sequences, which is obtained by comparing all sequences $s^{t1}_i$ of $t_1$ and $s^{t2}_i$ $t_2$ by index~$i$ (lines~\ref{alg:distance:l2}--\ref{alg:distance:l4}). The distance of two sequences is the difference of their lengths (line~\ref{alg:distance:l5}) increased by $1$ for each different event at index $j$ (lines~\ref{alg:distance:l6}--\ref{alg:distance:l9}). Thus, the distance is based on the differences of ordered events between the ordered sequences of two test~suites.
\end{inparaenum}

\begin{wrapfigure}[9]{r}{0.69\textwidth}\vspace{-5em}
\begin{center}
\resizebox{.69\textwidth}{!}{%
\begin{minipage}[T]{1\textwidth}%
\begin{algorithm}[H]
	\caption{$dist(t_1, t_2)$: compute distance between two test suites $t_1$ and $t_2$.}
	\label{alg:distance}
	\begin{algorithmic}[1]
		\Input Test suites $t_{1}$ and $t_{2}$, max. suite size $suite_{max}$, max. sequence length $seq_{max}$
		\Output Distance between $t_{1}$ and $t_{2}$
		\State \texttt{distance $\leftarrow 0$;}
		\For{\texttt{i $\leftarrow$ 0 to $suite_{max}$}} \Comment{iterate over all $suite_{max}$ test sequences} \label{alg:distance:l2}
		\State \texttt{$s^{t1}_i \leftarrow t_{1}[i]$;}   \Comment{$i^{th}$ test sequence of test suite $t_1$}
		\State \texttt{$s^{t2}_i \leftarrow t_{2}[i]$;} \Comment{$i^{th}$ test sequence of test suite $t_2$} \label{alg:distance:l4}
		\State \texttt{distance $\leftarrow$ distance + abs(|$s^{t1}_i$| - |$s^{t2}_i$|);} \Comment{length difference as\,distance} \label{alg:distance:l5}
		\For{\texttt{j $\leftarrow$ 0 to $seq_{max}$}}  \Comment{iterate over all $seq_{max}$ events} \label{alg:distance:l6}
		\If{\texttt{|$s^{t1}_i$| $\leq j$ \textbf{or} |$s1^{t2}_i$| $\leq j$}} break;
		\EndIf
		\If{\texttt{$s^{t1}_i$[j] $\neq$ $s^{t2}_i$[j]}} \Comment{event comparison by index $j$}
		\State \texttt{distance $\leftarrow$ distance + 1;} \Comment{events differ at index $j$ in both seqs.} \label{alg:distance:l9}
		\EndIf
		\EndFor
		\EndFor
		\State \textbf{return} \texttt{distance};
	\end{algorithmic}
\end{algorithm}
\end{minipage}
}
\end{center}
\end{wrapfigure}
This metric is motivated by the basic mutation operator of \Sapienz shuffling the order of test sequences within a suite, and the order of events within a sequence. It is common that the neighborhood relation is based on operators that make small changes to solutions~\cite{Moser+2017}.

\subsection{Experimental Setup}
\label{sec:fla-design}

To analyze the fitness landscape of \Sapienz, we extended \Sapienz with metrics that characterize the landscape.
We then executed \Sapienz on five apps, repeat each execution five times, and report mean values of the metrics for each app.\footnote{All experiments were run on single 4.0 Ghz quad-core PC with 16\,GB RAM, using 5 Android emulators (KitKat 4.4.2, API level 19) in parallel to test one app.}

The five apps we selected for the descriptive study are part of the 68 F-Droid benchmark apps~\cite{Choudhary+2015} used to evaluate \Sapienz~\cite{Mao+2016}.
We selected
\textit{\aarddict}, 
\textit{\munchLife}, 
and \textit{\passwordmanager}
since \Sapienz did not find any fault for these apps,
and \textit{\hotdeath}
and \textit{\kNineMail}\footnote{We used ver.\,5.207 of \kNineMail and not ver.\,3.512 as in the 68 F-Droid apps benchmark.}, for which \Sapienz did find faults~\cite{Mao+2016}.
Thus, we consider apps for which \Sapienz did and did not reveal crashes to obtain potentially different landscape characteristics that may present difficulties to~\Sapienz.

We configured \Sapienz as in~\cite{Mao+2016}. The crossover and mutation rates are set to 0.7 and 0.3 respectively. The population and offspring size is 50. An individual (test suite) contains 5 test sequences, each constrained to 20--500 events. Instead of 100 generations~\cite{Mao+2016}, we observed in initial experiments that the search stagnates earlier so that we set the number of generation to 40 (stopping criterion).

\subsection{Results}
\label{sec:fla-results}

\newcommand{\numberOfMetrics}{11}
\newcommand{\ppos}{{\textit{ppos}}\xspace}
\newcommand{\hv}{{\textit{hv}}\xspace}
\newcommand{\diam}{{\textit{diam}}\xspace}
\newcommand{\maxdiam}{{\textit{maxdiam}}\xspace}
\newcommand{\mindiam}{{\textit{mindiam}}\xspace}
\newcommand{\avgdiam}{{\textit{avgdiam}}\xspace}
\newcommand{\reldiam}{{\textit{reldiam}}\xspace}
\newcommand{\pconnec}{{\textit{pconnec}}\xspace}
\newcommand{\nconnec}{{\textit{nconnec}}\xspace}
\newcommand{\kconnec}{{\textit{kconnec}}\xspace}
\newcommand{\lconnec}{{\textit{lconnec}}\xspace}
\newcommand{\hvconnec}{{\textit{hvconnec}}\xspace}
\newcommand{\metric}[1]{\noindent$\bullet$\,\textit{#1}}
\newcommand{\metricgroup}[1]{\noindent\textbf{#1}~}

The results of our study provide an analysis of the fitness landscape of \Sapienz with respect to the global topology, particularly the diversity of solutions, how the solutions are spread in the search space, and evolve over time.
According to Smith~\etal~\cite[p.\,31]{SmithHLO02}, ``No single measure or description can possibly characterize any high-dimensional heterogeneous search space''. Thus, we selected $\numberOfMetrics$ metrics from literature and implemented them in \Sapienz, which characterize (1)~the Pareto-optimal solutions, (2)~the population, and (3)~the connectedness of Pareto-optimal solutions, all with a focus on diversity.
These metrics are computed after every generation so that we can analyze their development over~time.
In the following, we discuss these $\numberOfMetrics$ metrics and the results of the fitness landscape analysis. 
The results are shown in Figure~\ref{fig:fla} where the metrics (y-axis) are plotted over the 40 generations of the search (x-axis) for each of the five apps.

\input{fla}

\metricgroup{(1) Metrics for Pareto-Optimal Solutions}

\metric{Proportion of Pareto-optimal solutions (\ppos).}
For a population $P$, \ppos is the number of Pareto-optimal solutions $P_{opt}$ divided by the population size: $ppos(P) = \frac{|P_{opt}|}{|P|}$.
A high and especially strongly increasing \textit{ppos} may indicate that the search based on Pareto dominance stagnates due to missing selection pressure~\cite{Purshouse+Fleming2007}. A moderately increasing \ppos may indicate a successful~search.

For \Sapienz and all apps (see~Fig.~\ref{fig:ppos}), \ppos slightly fluctuates since a new solution can potentially dominate multiple previously non-dominated solutions. 
At the beginning of the search, \ppos is low (0.0--0.1), shows no improvement in the first 15--20 generations, and then increases for all apps except of \textit{\passwordmanager}.
Thus, the search seems to progress while the enormously increasing \ppos for \munchLife and \hotdeath might indicate a stagnation of the search.

\metric{Hypervolume (\hv).}
To further investigate the search progress, we compute the \hv after each generation. The \hv is the volume in the objective space covered by the Pareto-optimal solutions~\cite{Li+Yao2019,Wang+2016}.
Thus, an increasing \hv indicates that the search is able to find improved solutions, otherwise the \hv and search stagnate.

Based on the objectives of \Sapienz (max. crashes, max. coverage, and min. sequence length), we choose the nadir point ($0$~crashes, $0$~coverage, and sequence length of $500$) as the reference point for the \hv.
In Fig.~\ref{fig:hv}, the evolution of the \hv over time rather than the absolute numbers are relevant to analyze the search progress of \Sapienz.
While the \hv increases during the first 25 generations, it stagnates afterwards for all apps; for \kNineMail already after 5 generations. For \aarddict, \munchLife, and \hotdeath the \hv stagnates after the \ppos drastically increases (\cf~Fig.~\ref{fig:ppos}), further indicating a stagnation of the search.

\metricgroup{(2) Population-Based Metrics}

\metric{Population diameter (\diam).}
The \diam metrics measure the spread of all population members in the search space using a distance metric for individuals, in our case Algorithm~\ref{alg:distance}.
The maximum \diam computes the largest distance between any two individuals of the population $P$: $\maxdiam(P) = \max_{x_i, x_j \in P} dist(x_i, x_j)$ \cite{Bachelet99,Olorunda+Engelbrecht2008}, showing the absolute spread of $P$.
To respect outliers, we can compute the average \diam as the average of all pairwise distances between all individuals~\cite{Bachelet99}:
\vspace{-1em}
\begin{equation}
\avgdiam(P) = \frac{\sum_{i=0}^{|P|}\sum_{j=0,j \neq i}^{|P|}{dist(x_i, x_j)}}{ |P|(|P|-1)}
\label{eq:avgdiam}
\end{equation}
\vspace{-1.2em}

Additionally, we compute the minimum diameter to see how close individuals are in the search space, or even identical: $\mindiam(P) = \min_{x_i, x_j \in P} dist(x_i, x_j)$.

Concerning each plot for \Sapienz and all apps (see~Fig.~\ref{fig:diam}), the upper, middle, and lower curve are respectively \maxdiam, \avgdiam, and \mindiam.
For each curve, we see a clear trend that the metrics decrease over time, which is typical for genetic algorithms due to the crossover.
However, the metrics drastically decrease for \Sapienz in the first 25 generations. The \avgdiam decreases from $>$$1500$ to eventually $<$$200$ for each app. The \maxdiam decreases similarly but stays higher for \hotdeath and \kNineMail than for the other apps.
The development of the \avgdiam and \maxdiam indicates that all individuals are continuously getting closer to each other in the search space, thus becoming more similar.
The population even contains identical solutions as indicated by \mindiam reaching~$0$.

\metric{Relative population diameter (\reldiam).}
Bachelet~\cite{Bachelet99} further proposes the relative population diameter, which is the \avgdiam in proportion to the largest possible distance $d$: $\reldiam(P) = \frac{avgdiam(P)}{d}$.
This metric is indicative of the concentration of the population in the search space. A small \reldiam indicates that the population members are grouped together in a region of the space~\cite{Bachelet99}.

For \Sapienz, the largest possible distance $d$ between two test suites is 2500, in which case they differ in all events (up to 500 for a test sequence) for all of their five individual test sequences. For $d = 2500$ and all apps (\cf~Fig.~\ref{fig:reldiam}), \reldiam starts at a high level of around 0.9 indicating that the solutions are spread in the search space. Then, it decreases in the first 25 generations to around 0.4 (\aarddict, \munchLife, and \passwordmanager), and below 0.3 (\hotdeath and \kNineMail) indicating a grouping of the solutions in one or more regions of the~search~space.

\metricgroup{(3)\,Metrics Based on the Connectedness of Pareto-Optimal Solutions}

\noindent
The following metrics analyze the \textit{connectedness} and thus, clusters of Pareto-optimal solutions in the search space~\cite{Isermann77,Paquete09}.
For this purpose, we consider a graph in which Pareto-optimal solutions are vertices.
The edges connecting the vertices are labeled with weights $\delta$, which are the number of moves a neighborhood operator has to make to reach one vertice from another~\cite{Paquete09}. This results in a graph of fully connected Pareto-optimal solutions.
Introducing a limit $k$ on $\delta$ and removing the edges whose weights $\delta$ are larger than $k$ leads to varying sizes of connected components (clusters) in the graph.
This graph can be analyzed by metrics to characterize the Pareto-optimal solutions in the search space~\cite{Paquete09,Liefooghe14}.

In our case, the weights $\delta$ are determined by the distance metric for test suites based on the mutation operator of \Sapienz (\cf~Algorithm~\ref{alg:distance}).
We determined $k$ experimentally to be $300$ investigating values of $400$, $300$, $200$, and $100$. While a high value results in a single cluster of Pareto-optimal solutions, a low value results in a high number of singletons (\ie, clusters with one solution). Thus, two test suites (vertices) are connected (neighbors) in the graph if they differ in less than $300$ events across their test sequences as computed by Algorithm~\ref{alg:distance}.

\metric{Proportion of Pareto-optimal solutions in clusters (\pconnec).}
This metric divides the number of vertices (Pareto-optimal solutions) that are members of clusters (excl. singletons) by the total number of vertices in the graph~\cite{Paquete09}. A high \pconnec indicates a grouping of the Pareto-optimal solutions in the search~space. 

As shown in Fig.~\ref{fig:pconnec}, \pconnec is relatively low during the first generations before it increases for all apps. For \munchLife, \passwordmanager, and \hotdeath, \pconnec reaches 1 meaning that all Pareto-optimal solutions are in clusters, while it converges around 0.7 and 0.8 for \aarddict and \kNineMail respectively. This indicates that the Pareto-optimal solutions are grouped in the search~space.

\metric{Number of clusters (\nconnec).}
We further analyze in how many areas of the~search space (clusters) the Pareto-optimal solutions are grouped.
Thus, \nconnec counts the number of clusters in the graph~\cite{Paquete09,Liefooghe14}. A high (low) \nconnec indicates that the Pareto-optimal solutions are spread in many (few) areas of the search space.

Fig.~\ref{fig:nconnec} plots \nconnec for \Sapienz and all apps. The y-axis of each plot denoting \nconnec ranges from $0$ to $6$.
Initially, the Pareto-optimal solutions are distributed in 2--4 clusters, then grouped in $1$ cluster. An exception is \kNineMail for which there always exists more than $3$ clusters. Except for \kNineMail, this indicates that the Pareto-optimal solutions are grouped in one area of the search space. 

\metric{Minimum distance $k$ for a connected graph (\kconnec).}
This metric identifies $k$ so that all Pareto-optimal solutions are members of one cluster~\cite{Paquete09,Liefooghe14}.
Thus, \kconnec quantifies the spread of all Pareto-optimal solutions in the search space.

For \Sapienz, Fig.~\ref{fig:kconnec} plots \kconnec (ranging from $0$\,to\,$1400$) over the generations.
Similarly to the \diam metrics (\cf~Fig.~\ref{fig:diam}), \kconnec decreases, moderately for
\hotdeath~(from initially $\approx$700 to $\approx$600) and
\kNineMail ($\approx$1000\,$\rightarrow$\,$\approx$800), and drastically for
\passwordmanager ($\approx$1200\,$\rightarrow$\,$\approx$200),
\munchLife ($\approx$1000\,$\rightarrow$\,$\approx$200), and
\aarddict ($\approx$600\,$\rightarrow$\,$\approx$100).
This indicates that \textit{all} Pareto-optimal solutions are getting closer in the search space as the spread of the cluster is decreasing.

\metric{Number of Pareto-optimal solutions in the largest cluster (\lconnec).}
It determines the size of the largest cluster by the number of members~\cite{Liefooghe14},
showing how many Pareto-optimal solutions are in the most dense area of the search~space.

Fig.~\ref{fig:lconnec} plots \lconnec (ranging from $0$ to $50$ given the population size of $50$) over the generations.
\lconnec increases after $15$-$30$ generations to $20$ (\aarddict and \hotdeath) or even $50$ (\munchLife) solutions. This indicates that the largest cluster is indeed large so that many Pareto-optimal solutions are grouped in one area of the search space. In contrast, \lconnec stays always below $10$ indicating smaller largest clusters for \passwordmanager and \kNineMail than for the other apps.

\metric{Proportion of hypervolume covered by the largest cluster (\hvconnec).}
Besides \lconnec, we compute the relative size of the largest cluster in terms of hypervolume (\hv). Thus, \hvconnec is the proportion of the overall \hv covered by the Pareto-optimal solutions in the largest cluster. It quantifies how this cluster in the search space dominates in the objective space and contributes to the \hv.

For \Sapienz (\cf~Fig.~\ref{fig:hvconnec}), \hvconnec varies a lot during the first $10$ generations, then stabilizes at a high level for all apps.
For \aarddict, \munchLife, and \passwordmanager, the largest clusters covers $100\%$ of the \hv since there is only $1$ cluster left (\cf~\nconnec in Fig.~\ref{fig:nconnec}).
For \hotdeath, \hvconnec is close to $70\%$ indicating that there is $1$ other cluster covering $30\%$ of the \hv (\cf~\nconnec).
For \kNineMail, \hvconnec is around $90\%$ indicating that the other $2$--$3$ clusters (\cf~\nconnec) cover only $10\%$ of the \hv.
This indicates that the largest cluster covers the largest proportion of the \hv, and thus contributes most to the Pareto front.

\subsection{Discussion}
\label{sec:fla-discussion}

The results characterizing the fitness landscape of \Sapienz reveal insights about how \Sapienz manages the search problem of generating test suites for apps.

Firstly, the development of the proportion of Pareto-optimal solutions (\cf Fig.~\ref{fig:ppos}) and hypervolume (\cf~Fig.~\ref{fig:hv}) indicates a stagnation of the search after 25 generations. The drastically increasing proportion of Pareto-optimal solutions in some cases may indicate a problem of \textit{dominance resistance}, \ie, the search cannot produce new solutions that dominate the current, poorly performing but locally non-dominated solutions~\cite{Purshouse+Fleming2007}. 
In other cases, the proportion remains low, \ie, the search cannot find many non-dominated solutions.

Secondly, the development of the population diameters (\cf~Fig.~\ref{fig:diam}) indicate a decreasing diversity of \textit{all} solutions during the search. The development of the relative population diameter (\cf~Fig.~\ref{fig:reldiam}) witnesses this observation and indicates that the population members are concentrated in the search space~\cite{Bachelet99}.
The minimum diameter (\cf~Fig.~\ref{fig:diam}) even indicates that the population contains duplicates of solutions, which reduces the genetic variation in the population.

Thirdly, the development of the proportion of Pareto-optimal solutions in clusters (\cf~Fig.~\ref{fig:pconnec}) indicates a grouping of these solutions in the search space,
mostly in one cluster (\cf~Fig.~\ref{fig:nconnec}).
Another indicator for the decreasing diversity of the Pareto-optimal solutions is the decreasing minimum distance $k$ required to form one cluster of all these solutions (\cf~Fig.~\ref{fig:kconnec}).
Additionally, the largest cluster is often indeed large in terms of number of Pareto-optimal solutions (\cf~Fig.~\ref{fig:lconnec}), and hypervolume covered by these solutions (\cf~Fig.~\ref{fig:hvconnec}). Even if there exist multiple clusters of Pareto-optimal solutions, the largest cluster still contributes most to the overall hypervolume and thus, to the Pareto front.

In summary, the fitness landscape analysis of \Sapienz indicates a stagnation of the search while the diversity of all solutions decreases in the search space.

\section{\SapienzDiv}
\label{sec:sapienzdiv}

Given the fitness landscape analysis results, \Sapienz suffers from a decreasing diversity of solutions in the search space over time.
It is known that the performance of genetic algorithms is influenced by diversity~\cite{Crepinsek2013,PanichellaOPL15}. 
A low diversity may lead the search to a local optimum that cannot be escaped easily~\cite{Crepinsek2013}. 
Thus, diversity is important to address dominance resistance so that the search can produce new solutions that dominate poorly performing, locally non-dominated solutions~\cite{Purshouse+Fleming2007}. 
Moreover, Shir~\etal~\cite[p.\,95]{Shir09} report that promoting diversity in the search space does not hamper ``the convergence to a precise and diverse Pareto front approximation in the objective space of the original algorithm''.

Therefore, we extended \Sapienz to \SapienzDiv by integrating mechanisms into the search algorithm that promote the diversity of the population in the search space.\footnote{\SapienzDiv is available at: \url{https://github.com/thomas-vogel/sapienzdiv-ssbse19}.}	
We developed four mechanisms that extend the \Sapienz algorithm at different steps: at the initialization, before and after the variation, and at the selection. Algorithm~\ref{alg:sapienzdiv} shows the extended search algorithm of \SapienzDiv and highlights the novel mechanisms in blue.
We now discuss these mechanisms.

\newcommand{\mechanism}[1]{\noindent\textit{#1}}
\mechanism{Diverse initial population.}
As the initial population may effect the results of the search~\cite{Maaranen2006}, we assume that a diverse initial population could be a better start for the exploration.
Thus, we extend the generation of the initial population $P_{init}$ to promote diversity. Instead of generating $|P| = size_{pop}$ solutions, we generate $size_{init}$ solutions where $size_{init}>size_{pop}$ (line~\ref{initpop-start} in Algorithm~\ref{alg:sapienzdiv}). Then, we select those $size_{pop}$ solutions from $P_{init}$ that are most distant from each other using Algorithm~\ref{alg:distance}, to form the first population~$P$ (line~\ref{initpop-end}).

\input{sapienzdiv-algo}
\mechanism{Adaptive diversity control.}
This mechanism dynamically controls the diversity if the population members are becoming too close in the search space relative to the initial population. It further makes the algorithm adaptive as it uses feedback of the search to adapt the search (\cf~\cite{Crepinsek2013}).

To quantify the diversity $div_{pop}$ of population $P$, we use the average population diameter (\avgdiam) defined in Eq.~\ref{eq:avgdiam}. 
At the beginning of each generation, $div_{pop}$ is calculated  (line~\ref{calculate-div}) and compared to the diversity of the initial population $div_{init}$ (line~\ref{check-div}) calculated once in line~\ref{calculate-div-init}. The comparison checks whether $div_{pop}$ has decreased to less than $div_{limit} \times div_{init}$.
For example, the condition is satisfied for the given threshold $div_{limit}=0.4$ if $div_{pop}$ has decreased to less than $40\%$ of $div_{init}$.

In this case, the offspring $Q$ is obtained by generating new solutions using the original \Sapienz method to initialize a population (line~\ref{gen-offspring}). The next population is formed by selecting the $|P|$ most distant individuals from the current population $P$ and offspring $Q$ (line~\ref{selection-distance}). 
In the other case, the variation operators (crossover and mutation) of \Sapienz are applied to obtain the offspring (line~\ref{variation}) followed by the selection.
Thus, this mechanism promotes diversity by inserting new individuals to the population, having an effect of restarting the search.

\mechanism{Duplicate elimination.}
The fitness landscape analysis found duplicated test suites in the population. Eliminating duplicates is one technique to maintain diversity and improve search performance~\cite{Crepinsek2013,Ronald1998}.
Thus, we remove duplicates after reproduction and before selection in the current population and offspring (line~\ref{duplicate-elim}). Duplicated test suites are identified by a distance of $0$ computed by Algorithm~\ref{alg:distance}.

\mechanism{Hybrid selection.}
To promote diversity in the search space, the selection is extended by dividing it in two parts:
(1)~The non-dominated sorting of \nsgaTwo is performed as in \Sapienz (lines~\ref{nsga2sort-1}--\ref{nsga2sort-2} in Algorithm~\ref{alg:sapienzdiv}) to obtain the solutions $P'$ sorted by domination rank and crowding distance.
(2)~From $P'$, the best $(size_{pop} - n_{div})$ solutions form the next population $P$ where $size_{pop}$ is the size of $P$ and $n_{div}$ the configurable number of diverse solutions to be included in $P$ (line~\ref{takebest}). These $n_{div}$ diverse solutions $P_{div}$ are selected as the most distant solutions from the current population and offspring $PQ$ (line~\ref{selectdistant}) using the distance metric of Algorithm~\ref{alg:distance}. Finally, $P_{div}$ is added to the next population $P$ (line~\ref{newpopulation}).

While the \nsgaTwo sorting considers the diversity of solutions in the objective space (crowding distance), the selection of \SapienzDiv also considers the diversity of solutions in the search space, which makes the selection hybrid.

\section{Evaluation}
\label{sec:evaluation}

We evaluate \SapienzDiv in a head-to-head comparison with \Sapienz to investigate the benefits of the diversity-promoting mechanisms. Our evaluation targets five research questions (RQ) with two empirical studies similarly to~\cite{Mao+2016}:
\begin{itemize}
	\item[\textbf{RQ1}] How does the coverage achieved by \SapienzDiv compare to \Sapienz?
	\item[\textbf{RQ2}] How do the faults found by \SapienzDiv compare to \Sapienz?
	\item[\textbf{RQ3}] How does \SapienzDiv compare to \Sapienz concerning the length of their fault-revealing test sequences?
	\item[\textbf{RQ4}] How does the runtime overhead of \SapienzDiv compare to \Sapienz?
	\item[\textbf{RQ5}] How does the performance of \SapienzDiv compare to the performance of \Sapienz with inferential statistical testing?
\end{itemize}

\subsection{Experimental Setup}

We conduct two empirical studies, Study~1 to answer RQ1-4, and Study~2 to answer RQ5. The execution of both studies was distributed on eight servers\footnote{For each server: 2$\times$Intel(R) Xeon(R) CPU E5-2620 @ 2.00GHz, with 64GB RAM.} while each server runs one approach to test one app at a time using $10$ Android emulators (Android KitKat version, API~19).
We configured \Sapienz and \SapienzDiv as in the experiment for the fitness landscape analysis (\cf~Section~\ref{sec:fla-design}) and in~\cite{Mao+2016}.
The only difference is that we test each app for $10$ generations in contrast to Mao~\etal~\cite{Mao+2016} who test each app for one hour, since we were not in full control of the servers running in the cloud. However, we still report the execution times of both approaches (RQ4).
Moreover, we configured the novel parameters of \SapienzDiv as follows:
$size_{init}=100$,
$div_{limit}=0.5$, and
$n_{div}=15$.
For Study~1 we perform one run to test each app over $10$ generations by each approach. For Study~2 we perform $20$ repetitions of such runs for each app and approach.

\subsection{Results}

\noindent
\textbf{Study 1}~
In this study we use 66 of the 68 F-Droid benchmark apps\footnote{We exclude \textit{aGrep} and \textit{frozenbubble} as \Sapienz/\SapienzDiv cannot start these apps.} provided by Choudhary~\etal~\cite{Choudhary+2015} and used to evaluate \Sapienz~\cite{Mao+2016}.
The results on each app are shown in Table~\ref{tab:68} where
\textbf{S} refers to \Sapienz,
\textbf{Sd} to \SapienzDiv,
\textbf{Coverage} to the final statement coverage achieved,
\textbf{\#Crashes} to the number of revealed unique crashes,
\textbf{Length} to the average length of the minimal fault-revealing test sequences (or `--' if no fault has been found), and
\textbf{Time (min)} to the execution time in minutes of each approach to test the app over $10$ generations.

\begin{figure}[t]
	\centering
	\begin{minipage}[b]{0.29\textwidth}
		\centering
		\includegraphics[width=.95\textwidth]{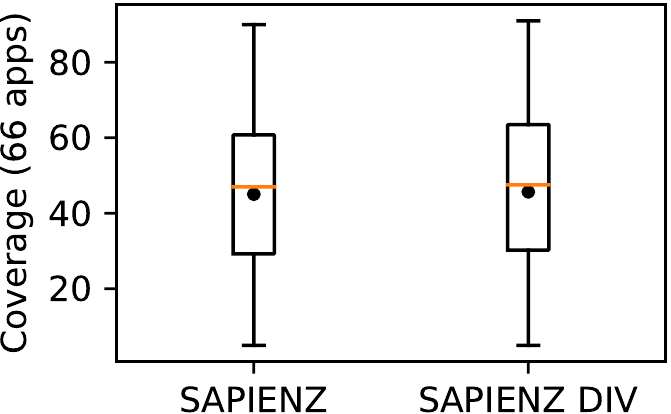}
		\vspace{-.5em}
		\caption{Coverage.}
		\label{fig:study1:coverage}
	\end{minipage}
	\begin{minipage}[b]{0.41\textwidth}
		\centering
		\captionsetup{type=table}
		\resizebox*{.99\textwidth}{!}{
		\begin{tabular}{lrr}
			\multicolumn{3}{c}{66 benchmark apps} \\
			\toprule
			& \Sapienz & \SapienzDiv \\
			\midrule
			\# App Crashed    & 43       & 46       \\
			\# Total Crashes  & 5974     & 6941     \\
			\# Unique Crashes & 119      & 141      \\
			\# Disjoint Crashes & 29    & 51		\\
			\# Intersecting Crashes & 90 & 90  	\\ \midrule
			Mean sequence length & 209  & 244  	   \\
			\bottomrule
		\end{tabular}
		}
		\vspace{-.5em}
		\caption{Crashes and seq. length.}
		\label{tab:study1:crashes}
	\end{minipage}%
	\begin{minipage}[b]{0.29\textwidth}
		\centering
		\includegraphics[width=.97\textwidth]{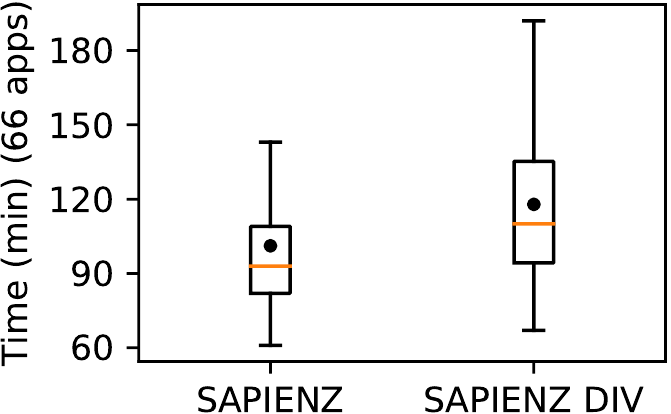}
		\vspace{-.5em}
		\caption{Time (min).}
		\label{fig:study1:time}
	\end{minipage}%
	\vspace{-2em}
\end{figure}

\renewcommand{\arraystretch}{.75}
\setlength{\tabcolsep}{4pt}
\begin{wraptable}[34]{r}{0.58\textwidth}
\vspace{-3em}
\caption{Results on the 66 benchmark apps.}
\vspace{-1em}
\label{tab:68}
\resizebox*{0.58\textwidth}{!}{\input{study1}}
\end{wraptable}
\textbf{RQ1} \Sapienz achieves a higher final coverage for 15 apps, \SapienzDiv for 24 apps, and both achieve the same coverage for 27 apps.
Fig.~\ref{fig:study1:coverage} shows that a similar coverage is achieved by both approaches on the 66 apps, in average $45.05$ by \Sapienz and $45.67$ by \SapienzDiv, providing initial evidence that \SapienzDiv and \Sapienz perform similarly with respect to coverage.

\textbf{RQ2} To report about the found faults, we count the total crashes, out of which we also identify the unique crashes (\ie, their stack traces are different from the traces of the other crashes of the app). Moreover, we exclude faults caused by the Android system (\eg, native crashes) and test harness (\eg, code instrumentation).

As shown in Table~\ref{tab:study1:crashes}, \SapienzDiv revealed more total (6941\,vs\,5974) and unique (141\,vs\,119) crashes, and found faults in more apps (46\,vs\,43) than \Sapienz.
Moreover,
it found 51 unique crashes undetected by \Sapienz,
\Sapienz found 29 unique crashes undetected by \SapienzDiv, and
both found the same 90 unique crashes.
The results for the 66 apps provide initial evidence that \SapienzDiv can outperform \Sapienz in revealing crashes.

\textbf{RQ3} Considering the minimal fault-revealing test sequences (\ie, the shortest of all sequences causing the same crash), their mean length is 244 for \SapienzDiv and 209 for \Sapienz on the 66 apps (\cf~Table~\ref{tab:study1:crashes}). This provides initial evidence that \SapienzDiv produces longer fault-revealing sequences than \Sapienz.

\textbf{RQ4} Considering the mean execution time of testing one app over 10 generation, \SapienzDiv takes 118 and \Sapienz 101 \textit{min.} for the 66 apps.
Fig.~\ref{fig:study1:time} shows that the diversity-promoting mechanisms of \SapienzDiv cause a noticeable runtime overhead compared to \Sapienz.
This provides initial evidence about the cost of promoting diversity at which an improved fault detection can be obtained.

\noindent
\textbf{Study 2}~
\begin{figure}[b!]
	\vspace{-1em}
	\centering
	\includegraphics[width=.96\textwidth]{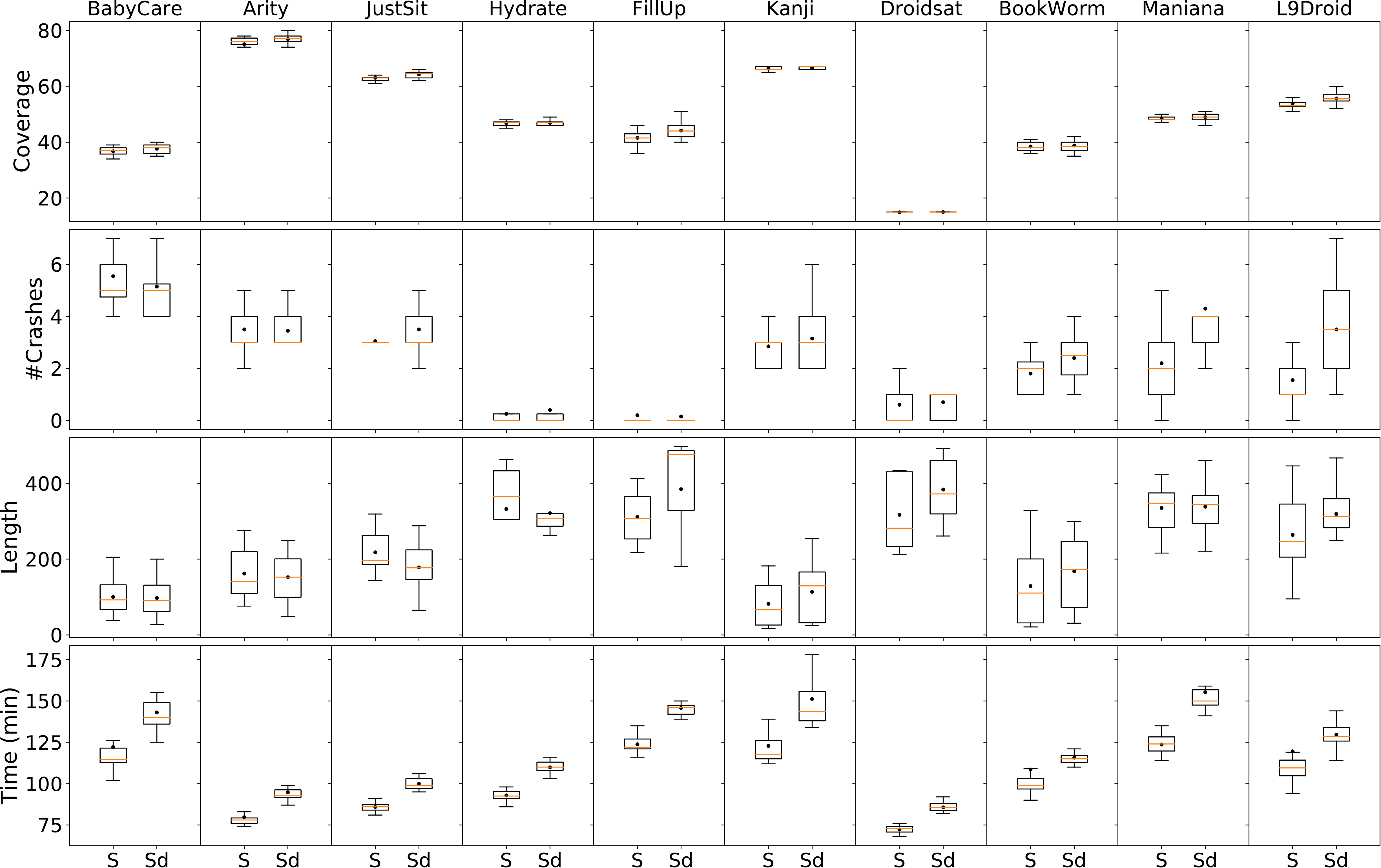}
	\vspace{-.75em}
	\caption{Performance comparison on $10$ apps for \SapienzDiv (Sd) and \Sapienz (S).}
	\label{fig:study2}
\end{figure}
In this study we use the same 10 F-Droid apps as in the statistical analysis in~\cite{Mao+2016}.
Assuming no Gaussian distribution of the results, we use the Kruskal-Wallis test to assess the statistical significance ($p$$<$0.05) and the Vargha-Delaney effect size $\hat{A}_{12}$ to characterize small, medium, and large differences between \SapienzDiv and \Sapienz ($\hat{A}_{12}>$ 0.56, 0.64, and 0.71 respectively).

\textbf{RQ5} The results are presented by boxplots in Fig.~\ref{fig:study2} for each of the 10 apps and concern: coverage, \#crashes, sequence length, and time (\cf~Study 1).
The $\hat{A}_{12}$ effect size for these concerns are shown in Table~\ref{tab:study2:effect}, which compares \SapienzDiv and \Sapienz (Sd-S) and emphasizes statistically significant results in bold.
\Sapienz significantly outperforms \SapienzDiv with large effect size on all apps for execution time. 
The remaining results are inconclusive.
\SapienzDiv significantly outperforms \Sapienz with large effect size on only 3/10 apps for coverage, 2/10 for \#crashes, and almost 1/10 for length. The remaining results are not statistically significant or do not indicate large differences.

\begin{table}[t]
\centering
\caption{Vargha-Delaney effect size (statistically significant results in bold).}
\label{tab:study2:effect}
\resizebox*{.79\textwidth}{!}{\input{effectsize}}
\end{table}

\subsection{Discussion}
Study~1 provided initial evidence that \SapienzDiv can find more faults than \Sapienz while achieving a similar coverage but using longer sequences. Especially,  the fault revelation capabilities of \SapienzDiv seemed promising, however, we could not confirm them by the statistical analysis in Study~2. The results of Study~2 are inconclusive in differentiating both approaches by their performance.
Potentially, the diversity promotion of \SapienzDiv does not results in the desired effect in the first 10 generations we considered in the studies. In contrast, it might show a stronger effect at later stages since we observed in the fitness landscape analysis that the search of \Sapienz stagnates after 25 generations.

\section{Threats to Validity}

\textbf{Internal validity.}
A threat to the internal validity is a bias in the selection of the apps we took from~\cite{Choudhary+2015,Mao+2016} although the 10 apps for Study~2 were selected by an ``unbiased random sampling''~\cite[p.\,103]{Mao+2016}.
We further use the default configuration of \Sapienz and \SapienzDiv without tuning the parameters to reduce the threat of overfitting to the given apps.
Finally, the correctness of the diversity-promoting mechanisms is a threat that we addressed by computing the fitness landscape analysis metrics with \SapienzDiv to confirm the improved diversity.

\noindent
\textbf{External validity.}
As we used 5 (for analyzing the fitness landscape) and 76 Android apps (for evaluating \SapienzDiv) out of over 2.500 apps on F-Droid and millions on Google Play, we cannot generalize our findings although we rely on the well-accepted ``68 F-Droid benchmark apps''~\cite{Choudhary+2015}.

\section{Related Work}
Related work exists in two main areas: approaches on test case generation for apps, and approaches on diversity in search-based software testing (SBST).

\noindent
{\bf Test case generation for apps.}
Such approaches use random, model-based, or systematic exploration strategies for the generation.
Random strategies implement UI-guided test input generators where events on the GUI are selected randomly~\cite{monkey}.
Dynodroid~\cite{Machiry2013} extends the random selection using weights and frequencies of events.
Model-based strategies such as PUMA~\cite{Hao2014}, DroidBot~\cite{Li2017}, MobiGUITAR~\cite{Amalfitano2015}, and Stoat~\cite{Su2017} apply model-based testing to apps. 
Systematic exploration strategies range from full-scale symbolic execution~\cite{Mirzaei2012} to evolutionary algorithms~\cite{Mahmood2014,Mao+2016}.
All of these approaches do not explicitly manage diversity, except of Stoat~\cite{Su2017} encoding diversity of sequences into the objective function.

\noindent
{\bf Diversity in SBST.}
Diversity of solutions has been researched for test case selection and generation.
For the former, promoting diversity can significantly improve the performance of state-of-the-art multi-objective genetic algorithms~\cite{PanichellaOPL15}.
For the latter, promoting diversity results in increased lengths of tests without improved coverage~\cite{Albunian2017}, matching our observation.
Both approaches witness that diversity promotion is crucial and its realization ``requires some care''~\cite[p.\,782]{Purshouse+Fleming2007}.

\section{Conclusions and Future Work}

In this paper, we reported on our descriptive study analyzing the fitness landscape of \Sapienz indicating a lack of diversity during the search. Therefore, we proposed \SapienzDiv that integrates four mechanisms to promote diversity. The results of the first empirical study on the 68 F-Droid benchmark apps were promising for \SapienzDiv but they could not be confirmed statistically by the inconclusive results of the second study with 10 further apps.
As future work, we plan to extend the evaluation to more generations to see the effect of \SapienzDiv when the search of \Sapienz stagnates. Moreover, we plan to identify diversity-promoting mechanisms that quickly yield benefits in the first few generations.

\vspace{1em}\noindent
\textbf{Acknowledgments}
This work has been developed in the \textit{FLASH} project (GR 3634/6-1) funded by the German Science~Foundation (DFG) and has been partially supported by the 2018 Facebook Testing and Verification research award.


\end{document}

%% file: fla.tex
\newcommand{\subfigwidth}{25mm}
\newcommand{\hsave}{\hspace*{-5mm}}
\newcommand{\mpagewidth}{30.1mm}

\newcommand{\metricsPath}{fig/metrics}
\newcommand{\figrow}[2]{
\hspace*{-2mm}
\subfigure[#1]{%
	\begin{minipage}[t]{\mpagewidth}
		\includegraphics[width=\subfigwidth]{\metricsPath/aarddict/#2}
	\end{minipage}\hsave
	\begin{minipage}[t]{\mpagewidth}
		\includegraphics[width=\subfigwidth]{\metricsPath/munchlife/#2}
	\end{minipage}\hsave
	\begin{minipage}[t]{\mpagewidth}
		\includegraphics[width=\subfigwidth]{\metricsPath/keepass/#2}
	\end{minipage}\hsave
	\begin{minipage}[t]{\mpagewidth}
		\includegraphics[width=\subfigwidth]{\metricsPath/hotdeath/#2}
	\end{minipage}\hsave
	\begin{minipage}[t]{\mpagewidth}
		\includegraphics[width=\subfigwidth]{\metricsPath/k9-mail/#2}
	\end{minipage}
}\vspace{-3mm}
}

\begin{figure}[hp]
\vspace{-1em}
\hspace*{-2mm}
\begin{minipage}[t]{\mpagewidth}
	\centering \hsave \textbf{\aarddict}
\end{minipage}\hsave
\begin{minipage}[t]{\mpagewidth}
	\centering \hsave \textbf{\munchLife}
\end{minipage}\hsave
\begin{minipage}[t]{\mpagewidth}
	\centering \hsave \textbf{passwordm.}
\end{minipage}\hsave
\begin{minipage}[t]{\mpagewidth}
	\centering \hsave \textbf{\hotdeath}
\end{minipage}\hsave
\begin{minipage}[t]{\mpagewidth}
	\centering \hsave \textbf{\kNineMail}
\end{minipage}%
\vspace{-.5em}

\figrow{Proportion of Pareto-optimal solutions (\ppos).\label{fig:ppos}}{proportion_pareto_optimal-crop}

\figrow{Hypervolume (\hv).\label{fig:hv}}{hv_hof-crop}

\figrow{Max., average, and min. population diameter (\maxdiam, \avgdiam, \mindiam).\label{fig:diam}}{population_diameter-crop}

\figrow{Relative population diameter (\reldiam).\label{fig:reldiam}}{rel_diameter-crop}

\figrow{Proportion of Pareto-optimal solutions in clusters (\pconnec).\label{fig:pconnec}}{pconnec_300-crop}

\figrow{Number of clusters (\nconnec).\label{fig:nconnec}}{nconnec_300-crop}

\figrow{Minimum distance $k$ for a connected graph (\kconnec).\label{fig:kconnec}}{kconnec-crop}

\figrow{Number of Pareto-optimal solutions in the largest cluster (\lconnec).\label{fig:lconnec}}{lconnec_300-crop}

\figrow{Proportion of hypervolume covered by the largest cluster (\hvconnec).\label{fig:hvconnec}}{hvconnec_300-crop}

\vspace{-.75em}
\caption{Fitness landscape analysis results for \Sapienz.}
\label{fig:fla}
\end{figure}

%% file: sapienzdiv-algo.tex
\algnewcommand{\LineComment}[1]{\State \(\triangleright\) #1}
\newcommand{\CodeColor}{\textcolor{blue}}

\begin{wrapfigure}[25]{r}{0.69\textwidth}\vspace{-5em}
\begin{center}
\resizebox{.69\textwidth}{!}{%
\begin{minipage}[T]{1\textwidth}%
\begin{algorithm}[H]
	\caption{Overall algorithm of \SapienzDiv}
	\label{alg:sapienzdiv}
	\begin{algorithmic}[1]
		\Input AUT $A$, 
		crossover probability $p$, 
		mutation probability $q$,
		max. generation $g_{max}$, 
		%
		population size $size_{pop}$,
		offspring size $size_{off}$, 
		\CodeColor{size of the large initial population $size_{init}$},
		\CodeColor{diversity threshold $div_{limit}$},
		\CodeColor{number of diverse solutions to include $n_{div}$}
		\Output UI model $M$, Pareto front $PF$, test reports $C$
		\State $M \leftarrow K_{0}; PF \leftarrow \emptyset; C \leftarrow \emptyset$; \Comment{initialization}
		\State \texttt{generation $g \leftarrow 0$;}
		\State \texttt{boot up devices $D$;} \Comment prepare devices/emulators that will run the app 
		\State \texttt{inject MOTIFCORE into $D$;} \Comment{install \Sapienz component for hybrid exploration}
		\State \texttt{static analysis on $A$;} \Comment{for seeding strings to be used for text fields of $A$}
		\State \texttt{instrument and install $A$;} \Comment{app under test is instrumented and installed on $D$}
		\State \CodeColor{\texttt{initialize population $P_{init}$ of size $size_{init}$;}} \Comment{\CodeColor{large initial population}}\label{initpop-start}
		\State \CodeColor{\texttt{$P = selectMostDistant(P_{init}, size_{pop})$;}} \Comment{\CodeColor{select $size_{pop}$ most distant individuals}}\label{initpop-end}
		\State \texttt{evaluate $P$ with MOTIFCORE and update $(M, PF, C)$;}
		\State \CodeColor{\texttt{$div_{init} = calculateDiversity(P)$;}} \Comment{\CodeColor{diversity of the initial population (Eq.~\ref{eq:avgdiam})}}\label{calculate-div-init}
		\While{$g < g_{max}$} 
		\State $g \leftarrow g + 1$;
		\State \CodeColor{$div_{pop} = calculateDiversity(P)$;} \Comment{\CodeColor{diversity of the current population (Eq.~\ref{eq:avgdiam})}} \label{calculate-div}
		\If {\CodeColor{$div_{pop} \leq div_{limit} \times div_{init}$}} \Comment{\CodeColor{check decrease of diversity}} \label{check-div}
		\State \CodeColor{\texttt{$Q \leftarrow$ generate offspring of size  $size_{off}$;}} \Comment{\CodeColor{$\approx$ generate a population}} \label{gen-offspring}
		\State \texttt{evaluate $Q$ with MOTIFCORE and update $(M,PF,C)$;}
		\State \CodeColor{\texttt{$P = selectMostDistant(P \cup Q, |P|)$;}} \Comment{\CodeColor{selection based on distance}} \label{selection-distance}
		\Else
		\State $Q \leftarrow wholeTestSuiteVariation(P, p, q);$ \Comment{create offspring} \label{variation}
		\State \texttt{evaluate $Q$ with MOTIFCORE and update $(M,PF,C)$;}
		\State \CodeColor{$PQ \leftarrow removeDuplicates(P \cup Q)$;} \Comment{\CodeColor{duplicate elimination}} \label{duplicate-elim}
		\State $\mathcal{F} \leftarrow sortNonDominated(PQ,|P|)$; \label{nsga2sort-1}
		\State $P' \leftarrow \emptyset$; \Comment{non-dominated individuals}
		\For{\textbf{each} front $F$ in $\mathcal{F}$}
		\If{$|P'| \geq  |P|$} break; \EndIf
		\State \texttt{$assignCrowdingDistance(F)$;}
		\For{\textbf{each} individual $f$ in $F$}
		\State $P' \leftarrow P' \cup f;$
		\EndFor
		\EndFor
		\State $P' \leftarrow sorted(P', \prec_{c});$ \label{nsga2sort-2}
		\State \CodeColor{$P \leftarrow P'[0:(size_{pop} - n_{div})]$;}     \Comment{\CodeColor{take best $(size_{pop} - n_{div})$ solution from $P'$}} \label{takebest}
		\State \CodeColor{$P_{div} = selectMostDistant(PQ, n_{div})$;} 		\Comment{\CodeColor{select $n_{div}$ most distant solutions}} \label{selectdistant}
		\State \CodeColor{$P = P \cup P_{div}$;} \Comment{\CodeColor{next population}} \label{newpopulation}
		\EndIf
		\EndWhile
		\State \textbf{return} $(M, PF, C)$;
	\end{algorithmic}
\end{algorithm}

\end{minipage}
}
\end{center}
\end{wrapfigure}

%% file: study1.tex
\begin{tabular}{|l|rr|rr|rr|rr|}
\toprule
\multirow{2}{*}{\textbf{Subject}} & \multicolumn{2}{c|}{\textbf{Coverage}} & \multicolumn{2}{c|}{\textbf{\#Crashes}} & \multicolumn{2}{c|}{\textbf{Length}} & \multicolumn{2}{c|}{\textbf{Time\,(min)}} \\ 
 & ~~~~\SapienzShort & \SapienzDivShort & ~~~~\SapienzShort & \SapienzDivShort & ~~~~\SapienzShort & \SapienzDivShort & ~~~~\SapienzShort & \SapienzDivShort\\ \midrule 
a2dp & 33 & 32 & 4 & 3 & 315 & 250 & 95 & 117 \\ 
aarddict & 14 & 14 & 1 & 1 & 103 & 454 & 69 & 74 \\ 
aLogCat & 66 & 67 & 0 & 2 & -- & 232 & 125 & 140 \\ 
Amazed & 69 & 69 & 2 & 1 & 193 & 69 & 67 & 78 \\ 
AnyCut & 64 & 64 & 2 & 0 & 244 & -- & 80 & 105 \\ 
baterrydog & 65 & 65 & 1 & 1 & 26 & 155 & 82 & 91 \\ 
swiftp & 13 & 13 & 0 & 0 & -- & -- & 88 & 105 \\ 
Book-Catalogue & 19 & 24 & 2 & 4 & 273 & 223 & 86 & 98 \\ 
bites & 33 & 35 & 1 & 1 & 76 & 39 & 78 & 91 \\ 
battery & 79 & 79 & 9 & 10 & 251 & 230 & 109 & 122 \\ 
addi & 19 & 18 & 1 & 1 & 39 & 31 & 87 & 133 \\ 
alarmclock & 62 & 62 & 6 & 9 & 133 & 279 & 143 & 163 \\ 
manpages & 69 & 69 & 0 & 0 & -- & -- & 81 & 92 \\ 
mileage & 34 & 33 & 5 & 6 & 252 & 286 & 100 & 114 \\ 
autoanswer & 16 & 16 & 0 & 0 & -- & -- & 78 & 90 \\ 
hndroid & 15 & 16 & 1 & 1 & 27 & 53 & 97 & 111 \\ 
multismssender & 57 & 54 & 0 & 0 & -- & -- & 88 & 102 \\ 
worldclock & 90 & 91 & 2 & 1 & 266 & 169 & 109 & 132 \\ 
Nectroid & 54 & 54 & 1 & 1 & 261 & 243 & 112 & 136 \\ 
acal & 21 & 20 & 7 & 7 & 222 & 187 & 140 & 160 \\ 
jamendo & 32 & 38 & 8 & 5 & 248 & 266 & 91 & 105 \\ 
aka & 45 & 44 & 8 & 9 & 234 & 226 & 140 & 171 \\ 
yahtzee & 47 & 47 & 1 & 1 & 356 & 215 & 79 & 86 \\ 
aagtl & 17 & 17 & 5 & 4 & 170 & 123 & 84 & 111 \\ 
CountdownTimer & 61 & 62 & 0 & 0 & -- & -- & 108 & 143 \\ 
sanity & 13 & 13 & 2 & 3 & 236 & 192 & 154 & 149 \\ 
dalvik-explorer & 69 & 69 & 2 & 4 & 148 & 272 & 143 & 162 \\ 
Mirrored & 42 & 44 & 10 & 9 & 114 & 179 & 219 & 245 \\ 
dialer2 & 41 & 41 & 2 & 0 & 223 & -- & 123 & 129 \\ 
DivideAndConquer & 79 & 81 & 3 & 3 & 75 & 55 & 90 & 94 \\ 
fileexplorer & 50 & 50 & 0 & 0 & -- & -- & 142 & 153 \\ 
gestures & 52 & 52 & 0 & 0 & -- & -- & 62 & 69 \\ 
hotdeath & 61 & 67 & 2 & 2 & 312 & 360 & 80 & 95 \\ 
adsdroid & 38 & 34 & 2 & 4 & 210 & 211 & 107 & 161 \\ 
myLock & 31 & 30 & 0 & 0 & -- & -- & 87 & 101 \\ 
lockpatterngenerator & 76 & 76 & 0 & 0 & -- & -- & 80 & 94 \\ 
mnv & 29 & 32 & 5 & 6 & 222 & 315 & 118 & 131 \\ 
k9mail & 5 & 6 & 1 & 2 & 445 & 412 & 93 & 113 \\ 
LolcatBuilder & 29 & 28 & 0 & 0 & -- & -- & 88 & 101 \\ 
MunchLife & 67 & 67 & 0 & 0 & -- & -- & 72 & 80 \\ 
MyExpenses & 45 & 41 & 2 & 3 & 359 & 309 & 115 & 133 \\ 
LNM & 57 & 58 & 1 & 1 & 292 & 209 & 104 & 120 \\ 
netcounter & 59 & 61 & 0 & 1 & -- & 256 & 95 & 106 \\ 
bomber & 72 & 71 & 0 & 0 & -- & -- & 63 & 72 \\ 
fantastischmemo & 25 & 28 & 3 & 6 & 325 & 275 & 86 & 96 \\ 
blokish & 49 & 62 & 2 & 2 & 197 & 204 & 75 & 86 \\ 
zooborns & 36 & 36 & 0 & 0 & -- & -- & 86 & 95 \\ 
importcontacts & 41 & 41 & 0 & 1 & -- & 462 & 94 & 106 \\ 
wikipedia & 26 & 31 & 1 & 3 & 95 & 373 & 69 & 88 \\ 
PasswordMaker & 50 & 49 & 1 & 2 & 86 & 216 & 103 & 112 \\ 
passwordmanager & 15 & 13 & 1 & 1 & 185 & 354 & 121 & 136 \\ 
Photostream & 30 & 31 & 2 & 3 & 195 & 161 & 143 & 192 \\ 
QuickSettings & 44 & 41 & 0 & 1 & -- & 307 & 96 & 130 \\ 
RandomMusicPlayer & 58 & 59 & 0 & 0 & -- & -- & 97 & 113 \\ 
Ringdroid & 40 & 23 & 2 & 4 & 126 & 208 & 280 & 188 \\ 
soundboard & 53 & 53 & 0 & 0 & -- & -- & 61 & 67 \\ 
SpriteMethodTest & 59 & 73 & 0 & 0 & -- & -- & 63 & 74 \\ 
SpriteText & 60 & 60 & 1 & 2 & 116 & 448 & 93 & 101 \\ 
SyncMyPix & 19 & 19 & 0 & 2 & -- & 402 & 97 & 143 \\ 
tippy & 70 & 72 & 1 & 1 & 384 & 459 & 84 & 105 \\ 
tomdroid & 50 & 52 & 1 & 1 & 152 & 90 & 93 & 111 \\ 
Translate & 48 & 48 & 0 & 0 & -- & -- & 82 & 99 \\ 
Triangle & 79 & 79 & 1 & 0 & 235 & -- & 93 & 89 \\ 
weight-chart & 47 & 49 & 3 & 4 & 171 & 283 & 88 & 109 \\ 
whohasmystuff & 60 & 66 & 0 & 1 & -- & 466 & 118 & 139 \\ 
Wordpress & 5 & 5 & 1 & 1 & 244 & 223 & 104 & 224 \\ 
\bottomrule
\end{tabular}

%% file: effectsize.tex
\begin{tabular}{lrrrrr}
\toprule
\textbf{App} & \textbf{Ver.} & \textbf{Coverage Sd-S} & \textbf{\#Crashes Sd-S} & \textbf{Length Sd-S} & \textbf{Time Sd-S} \\ 
\midrule
BabyCare & 1.5 & 0.66 & 0.46 & 0.52 & \textbf{0.15} \\ 
Arity & 1.27 & 0.67 & 0.49 & 0.54 & \textbf{0.05} \\ 
JustSit & 0.3.3 & \textbf{0.75} & \textbf{0.66} & \textbf{0.70} & \textbf{0.00} \\ 
Hydrate & 1.5 & 0.52 & 0.52 & 0.64 & \textbf{0.00} \\ 
FillUp & 1.7.2 & \textbf{0.77} & 0.47 & 0.33 & \textbf{0.00} \\ 
Kanji & 1.0 & 0.66 & 0.56 & 0.38 & \textbf{0.09} \\ 
Droidsat & 2.52 & 0.55 & 0.60 & 0.26 & \textbf{0.00} \\ 
BookWorm & 1.0.18 & 0.58 & 0.66 & 0.36 & \textbf{0.05} \\ 
Maniana & 1.26 & 0.66 & \textbf{0.82} & 0.49 & \textbf{0.00} \\ 
L9Droid & 0.6 & \textbf{0.75} & \textbf{0.81} & 0.32 & \textbf{0.11} \\ 
\bottomrule
\end{tabular}